\documentclass[12pt]{article}
\pdfoutput=1
\usepackage{putex}
\usepackage{graphicx}
\usepackage{caption}
\usepackage{subcaption}
\usepackage{epstopdf}
\usepackage{enumerate}
\usepackage{cite}
\usepackage{tensor}
\usepackage{slashed}
\usepackage[aligntableaux=center, boxsize=1em]{ytableau}
\usepackage[utf8]{inputenc}
\usepackage[
      colorlinks=true,
      linkcolor=blue,
      urlcolor=blue,
      filecolor=black,
      citecolor=red,
      ]{hyperref}
\usepackage{float}

\newcommand{\abs}[1]{\left\lvert #1 \right\rvert}

\newcommand {\be} {\begin {equation}}
\newcommand {\ee} {\end {equation}}

\newcommand {\bes} {\begin {equation*}}
\newcommand {\ees} {\end {equation*}}

\newcommand{\es}[2] {\begin{equation} \label{#1} \begin{split} #2 \end{split} \end{equation}}

\newcommand{\R}{\mathbb{R}}

\newcommand{\beq}{\begin{equation}}
\newcommand{\eeq}{\end{equation}}

\newcommand{\ov}{\over}

\def\le{\left}
\def\ri{\right}

\numberwithin{equation}{section}

\begin{document}

\institution{Exile}{Department of Particle Physics and Astrophysics, Weizmann Institute of Science, \cr Rehovot, Israel}

\title{Anomalous dimensions of monopole operators in scalar QED$_3$ with Chern-Simons term }

\authors{Shai M.~Chester\worksat{\Exile}}

\abstract{
We study monopole operators with the lowest possible topological charge $q=1/2$ at the infrared fixed point of scalar electrodynamics in $2+1$ dimension (scalar QED$_3$) with $N$ complex scalars and Chern-Simons coupling $|k|=N$. In the large $N$ expansion, monopole operators in this theory with spins $\ell<O(\sqrt{N})$ and associated flavor representations are expected to have the same scaling dimension to sub-leading order in $1/N$. We use the state-operator correspondence to calculate the scaling dimension to sub-leading order with the result $N-0.2789+O(1/N)$, which improves on existing leading order results. We also compute the $\ell^2/N$ term that breaks the degeneracy to sub-leading order for monopoles with spins $\ell=O(\sqrt{N})$.
}
\date{\today}

\maketitle

\tableofcontents

\section{Introduction}
\label{intro}

Monopole operators are defined in three dimensional Abelian gauge theories as local operators that are charged under the topological global symmetry $U(1)_\text{top}$ \cite{Polyakov:1975rs, Borokhov:2002ib}, whose conserved current and charge are
\es{topDef}{
j_\text{top}^\mu=\frac{1}{8\pi}\epsilon^{\mu\nu\rho}F_{\nu\rho}\,, \qquad q=\frac{1}{4\pi}\int_{\Sigma}F \,,
}
where $F_{\nu \rho} \equiv \partial_\nu A_\rho - \partial_\rho A_\nu$ is the gauge field strength, $\Sigma$ is a closed two-dimensional surface, and $j_\text{top}^\mu$ is conserved due to the Bianchi identity. In the normalization \eqref{topDef}, the charge $q$ is restricted by Dirac quantization to take the values $q\in\mathbb{Z}/2$.

Our goal in this paper is to compute the scaling dimensions of monopole operators with $q=1/2$ at the conformal fixed point of scalar QED$_3$ with $N$ flavors of complex scalars and nonzero Chern-Simons coupling $k$ to sub-leading order in $1/N$. The action for this theory can be written as \cite{Dyer:2015zha,Chester:2017vdh}\footnote{Note that we most closely follow the formulation in \cite{Dyer:2015zha}, which explicitly includes $\mathfrak{g}$ in the action, and is most convenient for subleading in $1/N$ calculations. In \cite{Chester:2017vdh}, the action is written in the more standard way without an explicit $\mathfrak{g}$, in which case explicit counterterms would have to be added when going to subleading order in $1/N$.}
\es{action}{
\mathcal{S}=N\int d^3x\left[\frac{1}{\mathfrak{g}}\left(\abs{(\nabla_\mu-iA_\mu)\phi^i}^2+i\lambda\left(|\phi^i|^2-1\right)\right)-\frac{i\kappa}{4\pi}\epsilon^{\eta\nu\rho}A_\eta\partial_\nu A_\rho\right]\,,
}
where $\phi^i$ are complex scalars in the fundamental representation of the flavor group $SU(N)$, $\lambda$ is a Lagrange multiplier field that imposes a length constraint, $\mathfrak{g}$ is a coupling constant, and we define $\kappa\equiv k/N$. When $\kappa=0$, this theory is equivalent to a non-linear sigma model with $\mathbb{CP}^{N-1}$ target space. The action \eqref{action} describes a conformal field theory (CFT) provided that we tune the coupling $\mathfrak{g}=\mathfrak{g}_c$ for some $\mathfrak{g}_c$. This CFT can be studied perturbatively in large $N$ and $k$ \cite{Appelquist:1981vg,Appelquist:1986fd,Appelquist:1988sr,Appelquist:1981sf}, where the fluctuations of $A_\mu$ and $\lambda$ are suppressed. 

Monopole operators in scalar QED$_3$ with $k\neq0$ play a central role in the recently discussed web of non-supersymmetric dualities \cite{Son:2015xqa,Aharony:2015mjs,Karch:2016sxi,Murugan:2016zal,Seiberg:2016gmd,Hsin:2016blu,Radicevic:2016wqn,Kachru:2016rui,Kachru:2016aon,Karch:2016aux,Metlitski:2016dht,Aharony:2016jvv,Benini:2017dus,Komargodski:2017keh}. For instance, when $N=k=1$ the lowest dimension monopole operator with $q=1/2$ is conjectured to be dual to a free fermion \cite{Seiberg:2016gmd}. For large $N$ and $k$, a more speculative version of this duality states that $q=1/2$ monopoles may be dual to baryons in a theory with $SU(N)_{N/2-1}$ gauge symmetry and $N$ fermions \cite{Komargodski:2017keh}. Monopole operators have also been used to access gauge theories using the conformal bootstrap \cite{Chester:2016wrc}, in which case large $N$ results are crucial to verify the accuracy of the non-perturbative bootstrap bounds.

As in \cite{Chester:2017vdh, Borokhov:2002ib, Pufu:2013vpa, Dyer:2013fja, Chester:2015wao, Metlitski:2008dw, Dyer:2015zha, Borokhov:2002cg}, we will compute the scaling dimension of the lowest dimension monopole operators using the state-operator correspondence, which identifies the scaling dimensions of monopole operators with charge $q$ with the energies of states in the Hilbert space on $S^2\times\mathbb{R}$ with $4\pi q$ magnetic flux through the $S^2$ \cite{Borokhov:2002ib}. The ground state energy on $S^2\times\mathbb{R}$ can then be computed in the large $N$ and $k$ limit using a saddle point expansion. When $k\neq0$, the Chern-Simons term induces a gauge charge proportional to $q$, so that the naive $S^2\times \mathbb{R}$ vacuum must be dressed by charged matter modes. Following \cite{Chester:2017vdh}, we can enforce this dressing by computing the small temperature limit of the thermal free energy on $S^2\times S^1_\beta$, where the radius $\beta$ of $S^1$ is related to the temperature $T$ by $T\equiv\frac1\beta$. The dressing is imposed by the saddle point value of the holonomy of the gauge field on $S^1_\beta$, which acts like a chemical potential for the matter fields. 

In \cite{Chester:2017vdh}, this method was used to compute the scaling dimension of the lowest dimension monopoles operators for all $q$ in scalar QED$_3$ with $k\neq0$ to leading order in $1/N$, as well as the finite temperature corrections to the thermal free energy to sub-leading order in $1/N$.\footnote{Monopole operators for other Abelian gauge theories with $k\neq0$ were also studied in \cite{Chester:2017vdh, Borokhov:2002ib}. For $k=0$ Abelian gauge theories, monopoles have been studied in fermionic theories \cite{Pufu:2013vpa, Dyer:2013fja, Chester:2015wao,PhysRevB.100.094443,Dupuis:2021yej}, scalar theories \cite{Murthy:1989ps,Metlitski:2008dw,Dyer:2015zha,2013PhRvL.111m7202B,2015arXiv150205128K}, and supersymmetric theories \cite{Borokhov:2002cg,Benini:2009qs,Benini:2011cma,Imamura:2011su,Kim:2009wb,Aharony:2015pla}.} These finite temperature corrections were used to show that monopole operators with many different spins and flavor symmetry representations had the same energy to sub-leading order in $1/N$. For instance, for $q=1/2$, monopoles with spin $\ell<O(\sqrt{N})$ and an associated spin-dependent flavor irrep are degenerate to sub-leading order. This degeneracy is broken by an energy splitting term $\delta E_\ell \propto\ell^2/N$, whose coefficient was not evaluated in \cite{Chester:2017vdh}. For general $\kappa$ and $q$, the leading order scaling dimension for these degenerate monopoles could be written as an infinite sum that could be evaluated numerically. For the special case $|\kappa|=q+1/2$, the expression takes the simple form
 \es{SpecialCase}{
  \abs{k} = \left( \abs{q} + \frac 12 \right) N : \qquad \qquad \Delta_q = \frac{2 \abs{q} (\abs{q}+1)(2\abs{q}+1)}{3} N + O(N^0) \,.
 }
In particular, for $q=1/2$ and $|\kappa|=1$, we have $\Delta_{1/2}=N+O(N^0)$, which has intriguing relations to the aforementioned dualities. For instance, for $N=k=1$ the scaling dimension $\Delta_{1/2}=1$ is predicted to be exact because the monopole is dual to a free fermion. For large $N$ and $k$, the leading order result would imply that the scaling dimension of the baryon in the dual $SU(N)_{N/2-1}$ gauge theory receives no quantum corrections. 

The results of this work can be seen as a completion of the scalar QED$_3$ analysis in \cite{Chester:2017vdh} for the case $q=1/2$. To compute the sub-leading correction to $\Delta_{1/2}$, given in \eqref{FINAL}, we compute the Gaussian fluctuations of the gauge and Lagrange multiplier fields around the saddle point. These results can also be used to compute the coefficient of $\delta E_\ell$, given in \eqref{splitFINAL}, which tells us how degenerate the lowest dimension monopoles are.

The rest of this paper is organized as follows. In Section \ref{setup}, we set up our computation. In Section \ref{leading}, we review the leading order analysis at large $N$. In Section \ref{subleading}, we compute the $1/N$ correction to $\Delta_{1/2}$. In Section \ref{splitting}, we compute the coefficient of $\delta E_\ell$. We end with concluding remarks in Section \ref{conc}. Several technical details of our computation are included in the Appendices.

\section{Setup}
\label{setup}

To study scalar QED$_3$ on $S^2\times {S^1_\beta}$ at large $N$, we rescale the fields in \eqref{action} and add a conformal mass term to get
\es{action2}{
\mathcal{S}=\int d^3x\left[\frac{\sqrt{g}}{\mathfrak{g}}\left(\abs{(\nabla_\mu-iA_\mu)\phi^i}^2+\frac{1}{4}|\phi^i|^2+i\lambda\left(|\phi^i|^2-N\right)\right)-N\frac{i\kappa}{4\pi}\epsilon^{\mu\nu\rho}A_\mu\partial_\nu A_\rho\right]\,,
}
where $\kappa\equiv k/N$ and $g$ is the determinant of the metric 
\es{metric}{
ds^2=d\theta^2+\sin^2\theta d\phi^2+d\tau^2\,,
}
where $\tau\in[-\beta/2,\beta/2)$ and $S^2$ has unit radius. We are interested in computing the thermal free energy ${F}_q$ in the presence of a magnetic flux $\int dA=4\pi q$ through $S^2$, with $q\in \mathbb{Z}/2$. After integrating out the matter in \eqref{action2}, we get the partition function
\es{free}{
Z_q=e^{-\beta F_q}=&\int_{\int_{S^2}F=4\pi q}DA\exp\left[-N\tr\log\left(-(\nabla_\mu-iA_\mu)^2+\frac14+i\lambda\right)\right.\\
&\left.\qquad\qquad\qquad\qquad+iN\int d^3x\left(\frac{\kappa}{4\pi}\epsilon^{\mu\nu\rho}A_\mu\partial_\nu A_\rho+\frac{\sqrt{g}}{\mathfrak{g}}\lambda\right)\right]\,.
}
We now expand $A_\mu$ and $\lambda$ around a saddle point by taking
\es{saddle}{
A_\mu=\mathcal{A}_\mu+a_\mu\,,\qquad i\lambda=\mu+i\sigma\,,
}
where $a_\mu$ and $\sigma$ are fluctuations around a background $A_\mu=\mathcal{A}_\mu$ and $i\lambda=\mu$\footnote{Note that \cite{Dyer:2015zha} defines $\mu^2_q=\mu$.} that satisfies the saddle-point conditions
\es{saddleCond}{
\frac{\delta F_q[A_\mu,\lambda]}{\delta A_\mu}\Bigg\vert_{\sigma=a_\mu=0}=\frac{\delta F_q[A_\mu,\lambda]}{\delta \lambda}\Bigg\vert_{\sigma=a_\mu=0}=0\,.
}
On $S^2\times S^1_\beta$ with magnetic flux $4\pi q$, the most general such background is $\mu$ constant and $\mathcal{A}_\mu^q$ given by
\es{background}{
\mathcal{A}_\tau=-i\alpha\,,\qquad \mathcal{F}_{\theta\phi}d\theta\wedge d\phi=q\sin\theta d\theta\wedge d\phi\,, 
}
where $\alpha=i\beta^{-1}\int_{S^1_\beta}A$ is a real constant called the holonomy of the gauge field. Physically, $\alpha$ corresponds to a chemical potential for the matter fields.

Since the integrand in \eqref{free} is proportional to $N$, the fluctuations $a_\mu$ and $\sigma$ have typical size of order $1/\sqrt{N}$, and so are suppressed at large $N$. The thermal free energy $F_q$ can then be expanded at large $N$ as
\es{largeN}{
F_q=NF_q^{(0)}+F_q^{(1)}+\frac1NF_q^{(2)}+\dots\,,
}
 where $ F_q^{(0)}$ comes from evaluating $F_q$ at the saddle point and $ F_q^{(1)}$  comes from the functional determinant of the fluctuations around the saddle point. These terms can furthermore be expanded at large $\beta$ to get
\es{largeB}{
F_q^{(0)}&=\Delta_q^{(0)}-\frac{1}{\beta}S^{(0)}_q+O(e^{-c\beta})\,,\\
F_q^{(1)}&=\Delta_q^{(1)}+\frac1\beta\left(\frac12\log N+\frac d2\log\beta+O(\beta^0)\right)\,,\\
}
for some integer $d$, where the temperature independent terms are identified with the scaling dimension, the $\beta^{-1}$ terms give the entropy of the degenerate monopole states, and the $\frac{\log\beta}{\beta}$ term is due to the $O(N^{-1})$ splitting of the degenerate monopole spectrum, which is a continuous spectrum at large $N$. In the following sections, we will mostly focus on $\Delta_q^{(0)}$ and $\Delta_q^{(1)}$. For more details on the temperature dependent terms, see \cite{Chester:2017vdh}. 

\section{Leading order free energy}
\label{leading}
We now briefly review the calculation of the leading order free energy $F_q=NF_q^{(0)}+O(N^0)$, for more details see \cite{Chester:2017vdh,Dyer:2015zha}. After setting $A_\mu$ and $\lambda$ to their saddle point values \eqref{background} in the free energy \eqref{free} and tuning the coupling $\mathfrak{g}$ to its critical value $\mathfrak{g}=\mathfrak{g}_c$, we find 
\es{leadF}{
F_q^{0}(\alpha,\mu)=\beta^{-1}\tr\log\left[-(\nabla_\mu-i\mathcal{A}_\mu)^2+\frac14+\mu\right]-2\kappa q\alpha-\frac{4\pi}{\mathfrak{g}_c}\mu\,.
}
The eigenvalues of the operator $\left[-(\nabla_\mu-i\mathcal{A}_\mu)^2+\frac14+\mu\right]$ on $S^2\times S^1_\beta$ with magnetic flux $4\pi q$ are $(\omega_n-\alpha)^2+\lambda_j^2$, where $\lambda_j$ are the energies of the modes of the theory quantized on $S^2\times \mathbb{R}$:
\es{spectrum}{
\lambda_j=\sqrt{(j+1/2)^2-q^2+\mu}\,,\qquad j\in\{q,\,q+1\,\dots\}\,,\qquad d_j=2j+1\,,
}
$d_j$ are the degeneracies of the modes, and we defined the Matsubara frequencies $\omega_n=\frac{2\pi n}{\beta}\,,n\in\mathbb{Z}$. Using this spectrum, we find
\es{free0}{
F_q^{(0)}(\alpha,\mu)=&-2\kappa q\alpha-\frac{4\pi}{\mathfrak{g}_c}\mu+\beta^{-1}\sum_{n\in\mathbb{Z}}\sum_{j\geq q}d_j\log\left[(\omega_n-i\alpha)^2+\lambda_j^2\right]\\
=&-2\kappa q\alpha-\frac{4\pi}{\mathfrak{g}_c}\mu+\beta^{-1}\sum_{j\geq q}d_j\log\left[2\left(\cosh(\beta\lambda_j)-\cosh(\beta\alpha)\right)\right]\,.
}
The last two terms are both divergent, but can be evaluated using zeta-function regularization. 

Lastly, we solve \eqref{saddleCond} for large $\beta$ to find the set of possible saddle points values of $\alpha$ and $\mu$. For $q=0$, it can be easily checked that $\alpha=\mu=0$ are saddles provided that ${\mathfrak{g}^{-1}_c}=O(N^0)$, so we can ignore ${\mathfrak{g}^{-1}_c}$ to leading order (see \cite{Dyer:2015zha} for more details). For $q>0$, the $\alpha$ saddle that gives the lowest real free energy is
\es{alphSad}{
\alpha(\kappa)=-\sgn(\kappa)\left(\lambda_q+\beta^{-1}\log\frac{\xi}{1+\xi}\right)+O(e^{-\beta(\lambda_{q+1}-\lambda_q)})\,,\qquad \xi\equiv\frac{2q|\kappa|}{d_q}\,.
}
We now plug this $\alpha$ into the $\mu$ saddle point equation and zeta-function regularize to get
\es{muSad}{
\sum_{j\geq q}\left(\frac{d_j}{2\lambda_j(\mu)}-1\right)-q+\frac{\xi d_q}{2\lambda_q(\mu)}=0\,.
}
We are interested in the special case $|\kappa|=q+1/2$, in which case we find the saddle $\mu=q^2$. Plugging these saddle point values back into \eqref{free0}, taking the large $\beta$ limit, and zeta-function regularizing, we get \eqref{SpecialCase}.

\section{Sub-leading order free energy}
\label{subleading}

We will now compute the sub-leading correction $F_q^{(1)}$ to the free energy. As shown in \eqref{largeB}, $F_q^{(1)}$ includes $\beta$-dependent terms, as well as the $\beta$-independent correction $\Delta_q^{(1)}$ to the scaling dimension. The $\beta$-dependent terms were already calculated in \cite{Chester:2017vdh}, so here we focus on $\Delta_q^{(1)}$.

To compute $F^{(1)}_q$, we expand \eqref{free} to quadratic order in the fluctuations $a_\mu$ and $\sigma$ around the saddle point values determined in section \ref{leading}.  The linear fluctuations vanish by definition of the saddle point, so we are left with a Gaussian integral:
\es{free1}{
\exp(-\beta F_q^{(1)})=&\int DaD\sigma\exp\biggl[-\frac{N}{2}\int d^3xd^3x'\sqrt{g}\sqrt{g'}\biggl(a_\mu(x)K^{\mu\nu}_q(x,x')a_\nu(x')\\
&\qquad\qquad\qquad\qquad+\sigma(x)K^{\sigma\sigma}_q(x,x')\sigma(x')+2\sigma(x)K^{\sigma\nu}_q(x,x')a_\nu(x')\biggr)\biggr]\,,
}
where
 \es{KernelsScalar}{
K_q^{\mu\nu}(x,x')&\equiv \frac{1}{N} \left[ -\frac{1}{\mathfrak{g}^2}\langle J^\mu(x)J^\nu(x')\rangle_{q}+\frac{2}{\mathfrak{g}}g^{\mu\nu}\delta(x-x')\langle J(x) \rangle_{q} \right] -{i\kappa\ov 2\pi}\,\delta(x,x') \,\epsilon^{\mu\nu\rho}\partial'_\rho \,,\\
K_q^{\sigma\nu}(x,x')&\equiv-\frac{i}{N\mathfrak{g}^2} \langle J(x) J^\nu(x')\rangle_{q}  \,,\qquad K_q^{\sigma\sigma}(x,x')\equiv \frac{1}{N\mathfrak{g}^2} \langle J(x) J(x')\rangle_{q}  \,, 
}
with 
\es{QED3J}{
J^\mu\equiv i\left[\phi^*_i\left(\nabla_\mu-i\mathcal{A}_\mu\right)\phi^i-\phi^i\left(\nabla_\mu+i\mathcal{A}_\mu\right)\phi^*_i   \right] \,, \qquad J \equiv \phi^*_i \phi^i \,.
}
The subscript $q$ on the expectation values denotes that they are computed under the assumption that $A_\mu$ and $\lambda$ are non-dynamical and fixed to their saddle point values $A_\mu=\mathcal{A}_\mu$ and $i\lambda=\mu$. These expectation values can be evaluated using Wick contractions in terms of the scalar thermal Green's function
\es{Green}{
\langle\phi^i(x)\phi^*_j(x')\rangle=\mathfrak{g}\delta^i_{j}G^q(x,x')
}
to get
\es{wickCPN}{
{ K}_{q,\text{mat}}^{\mu\nu}(x,x')&=D^\mu G_q(x,x')D^{\nu} G_q(x',x)-G_q(x',x)D^{\mu}D^{\nu} G_q(x,x')\\
&+D^\mu G_q(x',x)D^{\nu} G_q(x,x')-G_q(x,x')D^{\mu}D^{\nu} G_q(x',x) \\
&+2g^{\mu\nu}\delta(x-x')G_q(x,x)\,,\\
{ K}_{q,\text{CS}}^{\mu\nu}(x,x')&=-{i\kappa\ov 2\pi}\,\delta(x,x') \,\epsilon^{\mu\nu\rho}\partial'_\rho\,,\\
{ K}_q^{\sigma\nu}(x,x')&=G_q(x,x')D^{\nu}G_q(x',x)- G_q(x',x)D^{\nu}G_q(x,x') \,,\\
{ K}_q^{\sigma\sigma}(x,x')&=G_q(x,x')G_q(x',x) \,,
}
where $D^\mu=\partial^\mu-i\mathcal{A}_q^\mu(x)$ and $D^{\nu}=\partial'^{\nu}+i\mathcal{A}_q^{\nu}(x')$ denote the gauge-covariant derivatives in the presence of the background gauge fields, and we have separated the matter and Chern-Simons contributions to $K_q^{\mu\nu}(x,x')$. 

The path integral in \eqref{free1} is not yet well defined, because it contains many flat directions due to pure gauge modes. Since these pure gauge modes are independent of $q$, we can remove them by calculating $e^{-\beta F_q^{(1)}}/e^{-\beta F_0^{(1)}}$ and using the fact that $F_0^{(1)}=0$ because it corresponds to the scaling dimension of the unit operator. To perform these Gaussian path integrals, it is convenient to expand the fluctuations in spherical harmonics / Fourier modes:
 \es{aFourier}{
  a(x) &= \mathfrak{a}_{00}^\mathcal{E}(0) \frac{d\tau}{\sqrt{4 \pi \beta}}  + 
  \sum_{n=-\infty}^\infty \sum_{\ell=1}^\infty \sum_{m=-\ell}^\ell 
   \left[  \mathfrak{a}_{\ell m}^\mathcal{E}(\omega_n) {\cal E}_{n \ell m} (x) 
    + \mathfrak{a}_{\ell m}^\mathcal{B}(\omega_n) {\cal B}_{n \ell m} (x) \right]  \frac{e^{-i \omega_n \tau}}{\sqrt{\beta}}+ d \lambda(x)\,,\\
      \sigma(x) &=
  \sum_{n=-\infty}^\infty \sum_{\ell=0}^\infty \sum_{m=-\ell}^\ell 
    \mathfrak{b}_{\ell m}(\omega_n) Y_{\ell m} (\theta, \phi) \frac{e^{-i \omega_n \tau}}{\sqrt{\beta}} \,,
 }
where $d\lambda$ are pure gauge modes and ${\cal E}_{n \ell m} (x)$ and ${\cal B}_{n \ell m} (x)$, together with $d\tau/ (4 \pi \beta)$, form an orthonormal basis of polarizations for the one-form $a(x)$:
 \es{EBDef}{
    {\cal E}_{n \ell m}(x) = \frac{ \ell(\ell+1)  Y_{\ell m}d\tau  -i \omega_n dY_{\ell m}}{\sqrt{\ell(\ell+1)} \sqrt{\omega_n^2 + \ell(\ell+1)}}  \,, \qquad
     {\cal B}_{n \ell m}(x) = \frac{*_2 dY_{\ell m}}{\sqrt{\ell (\ell+1)}}  \,,
 }
where $*_2$ is the Hodge dual on $S^2$. From now on we will ignore the pure gauge modes $d\lambda$ because the integral over them cancels between the numerator and denominator in $e^{-\beta F_q^{(1)}}/e^{-\beta F_0^{(1)}}$. We can then define the Fourier transform of the kernels \eqref{KernelsScalar} as
\es{fourierKernscal}{
&\bold{K}_{q,\ell}(\omega_n)=\frac{4\pi}{2j+1}\int d^3x\sqrt{g} e^{i\omega_n(\tau-\tau')} \sum_{m=-\ell}^{\ell}  \\
& \begin{pmatrix} Y^{\dagger}_{\ell m}(x)K^{\sigma\sigma}_{q}(x,x')  {Y}_{\ell m}(x') & \mathcal{E}^{\dagger}_{\mu,n\ell m}(x)K^{\mu\sigma}_{q}(x,x')  Y_{\ell m}(x')   &   \mathcal{B}^{\dagger}_{\mu,n\ell m}(x)  K^{\mu\sigma}_{q}(x,x')  Y_{\ell m}(x') \\
Y^{\dagger}_{\ell m}(x)K^{\sigma\nu}_{q}(x,x')  \mathcal{E}_{\nu,n\ell m}(x') & \mathcal{E}^{\dagger}_{\mu,n\ell m}(x)K^{\mu\nu}_{q}(x,x')  \mathcal{E}_{\nu,n\ell m}(x')   &   \mathcal{B}^{\dagger}_{\mu,n\ell m}(x)  K^{\mu\nu}_{q}(x,x')  \mathcal{E}_{\nu,n\ell m}(x') \\
Y^{\dagger}_{\ell m}(x)K^{\sigma\nu}_{q}(x,x')  \mathcal{B}_{\nu,n\ell m}(x') &
\mathcal{E}^{\dagger}_{\mu,n\ell m}(x) K^{\mu\nu}_{q}(x,x')  \mathcal{B}_{\nu,n\ell m}(x')   &   \mathcal{B}^{\dagger}_{\mu,n\ell m}(x) K^{\mu\nu}_{q}(x,x')   \mathcal{B}_{\nu,n\ell m}(x') \end{pmatrix}\Bigg\vert_{x'=0},
}
where we used $S^2\times S^1_\beta$ symmetry to eliminate the integral over $x'$, and we will denote the entries of $\bold{K}_{q,\ell}(\omega_n)$ by $\sigma, \mathcal{E}, \mathcal{B}$, appropriately. For $\ell=0$, ${\cal B}_{n \ell m} (x)$ does not exist, so we should think of $\bold{K}_{q,0}(\omega_n)$ as a $2\times2$ matrix given by eliminating the $\mathcal{B}$ entries. If we furthermore restrict to $n\neq0$ for $\ell=0$, then we find that the $\mathcal{E}$ terms also vanish.\footnote{$dY_{00}$ does not exist and $Y_{00}d\tau$ is pure gauge on $S^2\times \mathbb{R}$\cite{Dyer:2015zha}, so on $S^2\times S^1_\beta$ the harmonic ${\cal E}_{n 0 0} (x)$ can only contribute to $\beta$-dependent terms, which come from $n=0$ \cite{Chester:2017vdh}.} This extra zero mode again cancels in $e^{-\beta F_q^{(1)}}/e^{-\beta F_0^{(1)}}$, so we can ignore it and in this case define $\bold{K}_{q,0}(\omega_n)\equiv K^{\sigma\sigma}_{q,0}(\omega_n) $ as a scalar. We can now plug in \eqref{aFourier} and \eqref{fourierKernscal} into the exponent of \eqref{free1} to get
 \es{NumExponentScal}{
  &{}-\frac{N}{2} \begin{pmatrix} b_{00}(0) \\ \mathfrak{a}_{00}^E(0)
   \end{pmatrix}^\dagger
   {\bf K}_{q, 0}(0) \begin{pmatrix} b_{00}(0) \\ \mathfrak{a}_{00}^E(0)
   \end{pmatrix}
    - \frac{N}{2} \sum_{n \neq 0} \abs{b_{00}(\omega_n)}^2 K^{\sigma\sigma}_{q, 0}(\omega_n) \\
  &{}-\frac{N}{2} \sum_{n=-\infty}^\infty \sum_{\ell=1}^\infty \sum_{m=-\ell}^\ell
   \begin{pmatrix} b_{\ell m}(\omega_n) \\ \mathfrak{a}_{\ell m}^E(\omega_n) \\ \mathfrak{a}_{\ell m}^B(\omega_n)
   \end{pmatrix}^\dagger
   {\bf K}_{q, \ell}(\omega_n) \begin{pmatrix} b_{\ell m}(\omega_n) \\ \mathfrak{a}_{\ell m}^E(\omega_n) \\ \mathfrak{a}_{\ell m}^B(\omega_n)
   \end{pmatrix} \,.
 }
 As shown in \cite{Chester:2017vdh}, the first term contributes purely to $\beta$-dependent terms, and so can be ignored for our purposes, while the other terms can be separated in the large $\beta$ limit into linear in $\beta$ terms and $\beta$-independent terms as
 \es{SmoothApprox}{
  {{\bf K}}_{q, \ell}(\omega_n) = \beta\overline{\bf K}_{q,\ell} \delta_{n0}+\widetilde{{\bf K}}_{q, \ell}(\omega)\big\vert_{\omega=\omega_n}+ O(e^{-(\lambda_{q+1}-\lambda_1) \beta}) \,,
}
where $\overline{\bf K}_{q,\ell} \delta_{n0}$ is a constant matrix, and the $\beta$-independent term $\widetilde{{\bf K}}_{q, \ell}(\omega)$ is approximated by a smooth function of $\omega \in \R$ with exponential precision. For $\ell=0$, we similarly define the scalar quantity $\widetilde{{ K}}^{\sigma\sigma}_{q, 0}(\omega)\big\vert_{\omega=\omega_n}$ as discussed above. 

We can now plug \eqref{NumExponentScal} into $e^{-\beta F_q^{(1)}}/e^{-\beta F_0^{(1)}}$, perform the Gaussian integrals, and take the large $\beta$ limit to get
\es{freeFinal}{
F_q^{(1)}=\int {d\omega\ov 2\pi }\left[\log \le({{\widetilde{ K}}^{\sigma\sigma}_{q, 0}(\omega) \ov {\widetilde{ K}}^{\sigma\sigma}_{0, 0} (\omega) }\ri)+ \sum_{\ell=1}^\infty  (2\ell+1)\log \det\le(\frac{\widetilde{\bf K}_{q, \ell,\text{mat}}(\omega)+\widetilde{\bf K}_{\ell,\text{CS}}(\omega) }{\widetilde{\bf K}_{0, \ell,\text{mat}}(\omega)+\widetilde{\bf K}_{\ell,\text{CS}}(\omega) }\ri)\right]+O(\beta^{-1})\,,
}
where for $\ell>0$ we have separated the matter and Chern-Simons contributions as in \eqref{wickCPN}, and the sum over $\omega_n$ has been converted into an integral to exponential precision in $\beta$. The quantity written in \eqref{freeFinal} is $\beta$-independent and so is identified with $\Delta_q^{(1)}$. For the $\beta$-dependent corrections, see \cite{Chester:2017vdh}. In the subsequent subsections, we will now explicitly compute the kernels appearing in \eqref{freeFinal}, and then perform the sum and integral to find $\Delta_q^{(1)}$. To ease the notation, we will no longer write $O(e^{-\beta})$ next to each expression.

\subsection{Chern-Simons kernel}
\label{CSkernel}

We begin by considering the Chern-Simons kernel $\widetilde{\bf K}_{\ell,\text{CS}}(\omega)$. As we see from \eqref{wickCPN}, $K_{q,\text{CS}}^{\mu\nu}(x,x')$ is local, and so can be computed without doing any integrals. By plugging $K_{q,\text{CS}}^{\mu\nu}(x,x')$ into \eqref{fourierKernscal} we find that the only nonzero terms are
\es{CSfourier}{
\bold{K}^{\mathcal{B}\mathcal{E}}_{q,\ell,\text{CS}}(\omega_n)=-\frac{2i\kappa}{2j+1}\sum_{m=-\ell}^\ell  \mathcal{B}^{\dagger}_{\mu,n\ell m}(0)\epsilon^{\mu\nu\rho}\partial'_\rho \left[  \mathcal{E}_{\nu,n\ell m}(x') e^{-i\omega_n \tau'} \right]\Big\vert_{x'=0}
}
and $\bold{K}^{\mathcal{E}\mathcal{B}}_{q,\ell,\text{CS}}(\omega_n)=\bold{K}^{\mathcal{B}\mathcal{E}}_{q,\ell,\text{CS}}(\omega_n)$. Since $\bold{K}^{\mathcal{B}\mathcal{E}}_{q,\ell,\text{CS}}(\omega_n)$ is manifestly $\beta$-independent, we can define $\widetilde{\bold{K}}^{\mathcal{B}\mathcal{E}}_{q,\ell,\text{CS}}(\omega)=\bold{K}^{\mathcal{B}\mathcal{E}}_{q,\ell,\text{CS}}(\omega_n)\vert_{\omega=\omega_n}$ by just extending $\omega_n$ to the real line. We then perform the sum in \eqref{CSfourier} to get
\es{CSfourier2}{
\widetilde{\bold{K}}^{\mathcal{B}\mathcal{E}}_{q,\ell,\text{CS}}(\omega)=\widetilde{\bold{K}}^{\mathcal{E}\mathcal{B}}_{q,\ell,\text{CS}}(\omega)=\frac{i\kappa}{2\pi}\sqrt{\omega^2+\ell(\ell+1)}\,.
}
In a pure Chern-Simons theory, the eigenvalues of $\bold{K}_{q,\ell}(\omega)$ would then be $\pm\frac{i\kappa}{2\pi}\sqrt{\omega^2+\ell(\ell+1)}$, which are the expected nonzero eigenvalues for the Chern-Simons operator $-\frac{i\kappa}{2\pi}*d$ on $S^2\times S^1_\beta$, as it squares to the vector Laplacian $-\frac{\kappa^2}{4\pi^2}*d*d$ that has the nonzero eigenvalue $-\frac{\kappa^2}{4\pi^2}\left(\omega^2+\ell(\ell+1)\right)$ \cite{Marino:2011nm}.

\subsection{Matter kernels at $q=0$}
\label{q0Kernel}

We next consider the matter kernels $\widetilde{\bf K}_{q,\ell,\text{mat}}(\omega)$, and start with the simpler case of $q=0$. As discussed in Section \ref{leading}, when $q=0$ we have $\alpha=\mu=0$, so in the large $\beta$ limit we expect to find the same answers in this case as the $S^2\times\mathbb{R}$ kernels computed in \cite{Dyer:2015zha}. 

In particular, the scalar thermal Green's function $G^0(x,x')$ on $S^2\times{S^1_\beta}$ should be the same as the Green's function on $S^2\times\mathbb{R}$ up to exponential precision in $\beta$. This latter Green's function can be found by conformally mapping the $\mathbb{R}^3$ expression $1/\left(4\pi|x-x'|\right)$, which yields
\es{freeG}{
G_0(x,x')=\frac{1}{4\pi\sqrt{2\left(\cosh\left(\tau-\tau'\right)-\cos\gamma\right)}}\,,
}
where $\gamma$ is the angle between the two points on $S^2$
\es{gamma}{
\cos\gamma=\cos\theta\cos\theta'+\sin\theta\sin\theta'\cos\left(\phi-\phi'\right)\,.
}

We can now plug $G_0(x,x')$ into \eqref{wickCPN}, take the Fourier transform \eqref{fourierKernscal} to compute ${\bf K}_{0,\ell,\text{mat}}(\omega_n)$, and send $\beta\to\infty$ to compute $\widetilde{\bf K}_{0,\ell,\text{mat}}(\omega)$. For instance, for ${K}^{\sigma\sigma}_{0,\ell}(\omega_n)$ we find
\es{q0D}{
{K}^{\sigma\sigma}_{0,\ell}(\omega_n)=\frac{1}{8\pi}\int_{-\beta/2}^{\beta/2} d\tau\int_0^\pi\sin\theta d\theta\frac{e^{i\omega_n\tau}P_\ell(\cos\theta)}{2(\cosh\tau-\cos\theta)}\,,
}
where the Legendre polynomial $P_\ell(\cos\theta)$ comes from the sum over $Y_{\ell m}(\theta,\phi)$ in \eqref{fourierKernscal}. We then perform the $\theta$ integral and expand at large $\tau$ to find
\es{q0D2}{
{K}^{\sigma\sigma}_{0,\ell}(\omega_n)=\frac{1}{8\pi}\int_{-\beta/2}^{\beta/2} d\tau\sum_{p=0}^\infty \frac{(p+\ell)!\Gamma(p+1/2)}{p!\Gamma(p+\ell+3/2)}e^{-(2p+\ell+1)|\tau|}e^{i\omega_n\tau} \,.
}
We can now explicitly take $\beta\to\infty$ and extend $\omega_n\to\omega\in\mathbb{R}$ to get $\widetilde{K}^{\sigma\sigma}_{0,\ell}(\omega)$ with exponential precision in $\beta$. The resulting $\beta$-independent integral was performed in \cite{Pufu:2013eda}, and yields
\es{KDfree}{
\widetilde{K}^{\sigma\sigma}_{0,\ell}(\omega)&\equiv D_\ell(\omega)=\left|\frac{\Gamma\left((\ell+1+i\omega)/2\right)}{4\Gamma\left((\ell+2+i\omega)/2\right)}\right|^2\,.
}
This expression will come up frequently, so we have defined the concise notation $D_\ell(\omega)$. Note that all the sums and integrals used to compute $\widetilde{K}^{\sigma\sigma}_{0,\ell}(\omega)$ were free of divergences.

If we write down expressions analogous to \eqref{q0D} for the other nonzero kernels, ${K}^{\mathcal{E}\mathcal{E}}_{0,\ell,\text{mat}}(\omega_n)$ and ${K}^{\mathcal{B}\mathcal{B}}_{0,\ell,\text{mat}}(\omega_n)$, we see that they suffer from two kinds of divergences. In \cite{Dyer:2015zha}, it was shown that these divergences can be uniquely regularized using gauge invariance and the $\mu$ saddle point condition. Here we will use a slicker method, whose ultimate justification lies in that it gives the same answer as \cite{Dyer:2015zha}.

The first divergence is due to the term $G_0(x,x)$ that appears in $K_{q,\text{mat}}^{\mu\nu}(x,x')$ as defined in \eqref{wickCPN}. From \eqref{freeG}, we see that $G_0(x,x)$ is polynomially divergent. To analyze this divergence, it is useful to rewrite $G_0(x,x')$ as an infinite sum \cite{Dyer:2015zha}:
\es{freeG2}{
G_0(x,x')=\frac{1}{4\pi}\sum_{j=0}^{\infty}\frac{2j+1}{2j+1}P_j(\cos\gamma)e^{-(j+1/2)|\tau-\tau'|}\,,
}
and then set $x'=x$ and use zeta-functions to find
\es{freeG3}{
G_0(x,x)=\frac{1}{4\pi}\sum_{j=0}^{\infty}\frac{2j+1}{2j+1}=\frac{\zeta(0,1/2)}{4\pi}=0\,.
}

The second divergence is due to the $\theta$ and $\tau$ integral in \eqref{fourierKernscal} over the other terms in $K_{q,\text{mat}}^{\mu\nu}(x,x')$, which diverges polynomially for $\tau=0$. We can regularize this by first performing the $\theta$ integral assuming $\tau\neq0$. The resulting function is exponentially damped in $\tau$, just as we saw explicitly in \eqref{q0D2}, so we can send $\beta\to\infty$ before performing the $\tau$ integral to exponential precision in $\beta$. Finally, we perform the $\tau$ integral by deforming the contour around $\tau=0$, which yields the finite results
\es{q0D2other}{
\widetilde{K}^{\mathcal{B}\mathcal{B}}_{0,\ell,\text{mat}}(\omega)&=\frac{\omega^2+\ell^2}{2}D_{\ell-1}(\omega)\,,\\
\widetilde{K}^{\mathcal{E}\mathcal{E}}_{0,\ell,\text{mat}}(\omega)&=\frac{\omega^2+\ell(\ell+1)}{2}D_{\ell}(\omega)\,.
}
The expressions \eqref{q0D2other} match those in \cite{Dyer:2015zha}, which justifies our choice of regularization.\footnote{\cite{Dyer:2015zha} uses a different vector basis to compute the kernels, for the relation to our basis see Appendix \ref{conventions}.}

\subsection{Scalar thermal Green's function}
\label{green}

We would next like to compute the matter kernels for $q\neq0$, for which we will need the scalar thermal Green's function $G^q(x,x')$ in the large $\beta$ limit. We will focus on the special case $|\kappa|=q+1/2$, where we will find that $G^q(x,x')$ can be written in closed form.

We can define $G^q(x,x')$ by
\es{GreenDef}{
\left[-\left(\nabla_\mu-i\mathcal{A}_\mu\right)^2+\frac14+\mu\right]G_q(x,x')=\delta(x-x')\,.
}
In \cite{Chester:2017vdh}, this differential equation was solved for general $q$ and $\kappa$ to find
\es{GreenDef1}{
G_q(x,x')&=e^{\alpha(\tau-\tau')}\sum_{j,m}\left[\frac{Y_{q,jm}(\theta,\phi)Y^*_{q,jm}(\theta',\phi')}{2\lambda_j}  \left(e^{-\lambda_j|\tau-\tau'|}+\frac{e^{-\lambda_j(\tau-\tau')}}{e^{\beta(\lambda_j-\alpha)}-1}+\frac{e^{\lambda_j(\tau-\tau')}}{e^{\beta(\lambda_j+\alpha)}-1}\right) \right]\,, \\
}
where $Y_{q,jm}(\theta,\phi)$ are the monopole spherical harmonics introduced in \cite{Wu:1976ge,Wu:1977qk}. If we now plug in the value of $\alpha$ given in \eqref{alphSad}, then to exponential precision in large $\beta$ we get
\es{GreenDef2}{
G_q(x,x')&=e^{\alpha(\tau-\tau')}e^{-2iq\Theta}\left[\frac{ q|\kappa|}{(2q+1)\lambda_q} e^{\sgn(\kappa)\lambda_q(\tau-\tau')}F_{q,q}(\gamma) +\sum_{j\geq q}\frac{e^{-\lambda_j|\tau-\tau'|}}{2\lambda_j} F_{q,j}(\gamma)\right] \,,
}
where in the second line we performed the sum over $m$ following \cite{Dyer:2015zha} using the definitions
\es{Fpoly}{
F_{q,j}(\gamma)\equiv&\sqrt{\frac{2j+1}{4\pi}}Y_{q,j(-q)}(\gamma,0)\,,\\
e^{i\Theta}\cos(\gamma/2)\equiv&\cos(\theta/2)\cos(\theta'/2)+e^{-i(\phi-\phi')}\sin(\theta/2)\sin(\theta'/2)\,.
}
Recall from Section \ref{leading} that for the special case $|\kappa|=q+1/2$ we find the saddle $\mu=q^2$, so that from \eqref{spectrum} we get $\lambda_j=j+1/2$. Using this relation, along with the identity
\es{Fident}{
F_{q,j}(\cos\gamma)=\frac{2j+1}{4\pi}\left(\frac{1+\cos\gamma}{2}\right)^qP_{j-q}^{0,2q}(\cos\gamma)\,,
}
we can write the Green's function for $|\kappa|=q+1/2$ as 
\es{GreenDef3}{
G_{|\kappa|-1/2}(x,x')=\frac{e^{\alpha(\tau-\tau')}}{4\pi}&\left(\frac{1+\cos\gamma}{2}\right)^qe^{-2iq\Theta}\left[qe^{(q+1/2)(\tau-\tau')\sgn{\kappa}} \right.\\
&\left.+\sum_{j\geq q}{e^{-(j+1/2)|\tau-\tau'|}} P_{j-q}^{(0,2q)}(\cos\gamma)\right] \,.
}
Finally, we sum the Jacobi polynomials $P_{j-q}^{(0,2q)}$ using their generating function
\es{jacobi}{
\sum_{n=0}^\infty P_n^{a,b}(x)z^n&=\frac{2^{a+b}}{z^{(a+b+1)/2}}s^{-1}\left(z^{-1/2}-z^{1/2}+s\right)^{-a}\left(z^{-1/2}+z^{1/2}+s\right)^{-b}\,,\\
s&\equiv\sqrt{z+z^{-1}-2x}\,,
}
to find
\es{GreenDef4}{
G_{|\kappa|-1/2}(x,x')&=\frac{e^{\alpha(\tau-\tau')}}{4\pi}\left[\frac{1}{ s}\left(\frac{2e^{-i\Theta}\cos\frac\gamma2}{2\cosh\frac{\tau-\tau'}{2}+ s }\right)^{2q} +qe^{(q+1/2)(\tau-\tau')\sgn{\kappa}}\left( e^{-i\Theta}\cos\frac\gamma2 \right)^{2q} \right]\,,\\
s&=\sqrt{2\cosh(\tau-\tau')-2\cos\gamma}\,.
}

\subsection{Matter kernels at $q=1/2$}
\label{matter}
We will now compute the matter contributions to $\widetilde{\bold K}_{q,\ell}(\omega)$ for the specific case $q=1/2$ and $|\kappa|=1$. 

We find it convenient to separate the Green's function \eqref{GreenDef4} in this case as
\es{GreenFinal}{
G_{1/2}(x,x')&=e^{\alpha(\tau-\tau')}\left( G(x,x')+e^{(\tau-\tau')\sgn\kappa}\hat G(x,x') \right)\,,\\
G(x,x')&\equiv\frac{1}{4\pi \sqrt{2\cosh(\tau-\tau')-2\cos\gamma}}\left[\frac{2e^{-i\Theta}\cos\frac\gamma2}{2\cosh\frac{\tau-\tau'}{2}+ \sqrt{2\cosh(\tau-\tau')-2\cos\gamma} }\right]\,,\\
\hat G(x,x')&\equiv \frac{e^{-i\Theta}\cos\frac\gamma2 }{8\pi}\,,
}
where $G(x,x')$ is the same as the $\kappa=0$ Green's function computed in \cite{Dyer:2015zha} for $\mu=q^2$, while $\hat G(x,x')$ is a new contribution due to the nontrivial $\alpha$ that only appears because $k\neq0$. In the microscopic description, $ \hat G(x,x')$ comes from the modes used to dress the bare monopole in order to cancel the gauge charge induced by $k$. Both $ \hat G(x,x')$ and $ G(x,x')$ are invariant under the $CP$ transformation $\tau-\tau'\to\tau'-\tau$, but $ \hat G(x,x')$ is multiplied by a phase that violates $CP$. 

All the terms that appear in the matter kernels \eqref{wickCPN} with $q=1/2$ either contain two Green's functions $G_{1/2}(x,x')$ and $G_{1/2}(x',x)$, or $G_{1/2}(x,x)\delta(x,x')$, so the overall phase $e^{\alpha(\tau-\tau')}$ can be neglected as long as we replace the covariant derivative $D_\mu$ by $\hat D_\mu\equiv D_\mu\vert_{(\nabla_\tau-\alpha)\to\nabla_\tau}$. The resulting terms can then be divided into three categories. Firstly, there are terms given by setting $G_q(x,x')\to \hat G(x,x')$ in \eqref{wickCPN}, and ignoring the $G_q(x,x)$ term. These terms are independent of $\tau$, and so only contribute to the linear in $\beta$ term $\overline{\bf K}_{q,\ell} $ defined in \eqref{SmoothApprox}, so we will not consider them. Secondly, there are terms given by setting $G_q(x,x')\to G(x,x')$ in \eqref{wickCPN}, and so are $CP$-preserving terms. Lastly, the remaining terms in $\eqref{wickCPN}$ receive contribution from pairs of $ G(x,x')$ and $\hat G(x',x)$ or just a single $\hat G(x',x)$, and so are $CP$-violating terms. In the next couple subsections, we will compute these latter two categories separately.

\subsubsection{$CP$-preserving terms}
\label{CP}

We will now compute the $CP$-preserving kernels $\widetilde{\bold K}_{1/2,\ell,CP}(\omega)$. Just as we saw with the $q=0$ kernel in Section \ref{q0Kernel}, the function of $\tau$ that we get after we perform the $\theta$ integral in \eqref{fourierKernscal} is exponentially damped in $\tau$, so we can send $\beta\to\infty$ before we do the $\tau$ integral in \eqref{fourierKernscal} with exponentially precision in $\beta$.

As an illustrative example, let us begin by computing the $CP$-preserving kernel $\widetilde{ K}^{\sigma\sigma}_{1/2,\ell,CP}(\omega)$. Replacing $G(x,x')$ in \eqref{wickCPN} by $G(x,x')$ given in \eqref{GreenFinal} and plugging into the $\beta\to\infty$ limit of \eqref{fourierKernscal} we get
\es{DposCP}{
\widetilde{K}^{\sigma\sigma}_{1/2,\ell,CP}(\omega)=&\frac{1}{16\pi}\int_{-\infty}^{\infty} d\tau\int_0^\pi\sin\theta d\theta e^{i\omega\tau}P_\ell(\cos\theta)\left[ \frac{1}{(\cosh\tau-\cos\theta)}\right.\\
&\left.\qquad\qquad\qquad\qquad\qquad\qquad\qquad\qquad+\frac{\sqrt{2\cosh\tau-2\cos\theta}-2\cosh\frac\tau2}{(1+\cos\theta)\sqrt{2\cosh\tau-2\cos\theta}}\right]\,.
}
We already encountered the first term in the brackets in \eqref{q0D}, so its Fourier transform is $D_\ell(\omega)$ as defined in \eqref{KDfree}. For the second term, we can expand at large $\tau$ and perform the $\theta$ integral term by term to get
\es{DposCP}{
\frac{(-1)^{\ell}}{4\pi}\int_{-\infty}^\infty d\tau e^{i\omega\tau}\sum_{p=\ell+1}^\infty \frac{(-1)^p}{p}e^{-p|\tau|}\,,
}
and then perform the Fourier transform to get the final answer
\es{CPD}{
\widetilde{K}^{\sigma\sigma}_{1/2,\ell,CP}(\omega)=&D_\ell(\omega)+\frac{1}{2\pi}\sum_{p=\ell+1}^\infty\frac{(-1)^{p+\ell}}{\omega^2+p^2}\,.
}
As a consistency check, if we send $\ell\to\infty$ we recover the $q=0$ kernel $\widetilde{K}^{\sigma\sigma}_{0,\ell}(\omega)=D_\ell(\omega)$. Note that all the sums and integrals in this computation were finite. We can perform a similar procedure for $\widetilde{K}^{\sigma\mathcal{B}}_{1/2,\ell,CP}(\omega)$, and get a finite but more complicated expression that we give in Appendix \ref{kernels}. 

The other nonzero $CP$-preserving kernels are  $\widetilde{K}^{\mathcal{E}\mathcal{E}}_{1/2,\ell,CP}(\omega)$ and  $\widetilde{K}^{\mathcal{B}\mathcal{B}}_{1/2,\ell,CP}(\omega)$, and they suffer from the same two kinds of polynomial divergences that we saw for $q=0$ in Section \ref{q0Kernel}: the first from the term $G{(x,x)}$, the second from the $\theta$ and $\tau$ integral in \eqref{fourierKernscal} when $\tau=0$. We will regularize both just as we did in Section \ref{q0Kernel}, and justify our regularization by comparing to the results of \cite{Dyer:2015zha}.

To regularize the first divergence, we write $G(x,x)$ as the infinite sum given in \eqref{GreenDef2} and use zeta-functions:
\es{Gdiv}{
G(x,x)=\frac{1}{4\pi}\sum_{j=1/2}^{\infty}\frac{2j+1}{2j+1}=\frac{\zeta(0)}{4\pi}=-\frac{1}{8\pi}\,.
}
To regularize the second divergence, we first perform the $\theta$ integral assuming $\tau\neq0$, and then the $\tau$ integral by deforming the contour around $\tau=0$. The resulting expressions are also quite complicated, and are given in Appendix \ref{kernels}. Since these $CP$ preserving kernels were computed using the $\kappa=0$ Green's function, we can compare to the $\kappa=0$ results of \cite{Dyer:2015zha}. As we show in Appendix \ref{conventions}, our results match those of \cite{Dyer:2015zha}, which justifies our choice of regularization.

\subsubsection{$CP$-violating terms}
\label{noCP}

Next, we calculate the $CP$-violating kernels $\widetilde{\bold K}_{1/2,\ell,NCP}(\omega)$. These kernels differ from the $CP$-preserving kernels $\widetilde{\bold K}_{1/2,\ell,CP}(\omega)$ in two important ways. 

Firstly, $\widetilde{\bold K}_{1/2,\ell,CP}(\omega)$ receives contributions from the Green's function $\hat G(x,x)$ that only exists for $\kappa\neq0$, so we cannot compare our results to the $\kappa=0$ results of \cite{Dyer:2015zha}. On the other hand, the entire computation of $\widetilde{\bold K}_{1/2,\ell,CP}(\omega)$ is finite so there is no regularization ambiguity. In particular, from \eqref{GreenFinal} we see that $\hat G(x,x)=\frac{1}{8\pi}$, which exactly cancels the regularized value $ G(x,x)=-\frac{1}{8\pi}$ so that the total Green's function $G_{1/2}(x,x)$ vanishes, just as we found for $q=0$ in \eqref{freeG3}, and as was found for $\kappa=0$ and $q\neq0$ in \cite{Dyer:2015zha}.

Secondly, the position space kernels now include both terms that are exponentially damped in $\tau$, as well as terms that are independent of $\tau$. If we perform the $\tau$ integral in \eqref{fourierKernscal} and then send $\beta\to\infty$, the former terms contribute to $\widetilde{\bold K}_{1/2,\ell,NCP}(\omega)$ while the latter terms are linear in $\beta$ and so contribute to $\overline{\bold K}_{1/2,\ell}(\omega_n)$, as defined in \eqref{SmoothApprox}. If we send $\beta\to\infty$ before we compute the $\tau$ integral, then these linear in $\beta$ terms will appear as delta functions, according to the identity
\es{delta}{
2\pi \delta(\omega)=\int_{-\infty}^\infty e^{i\omega \tau}d\tau=\lim_{\beta\to\infty} \int_{-\beta/2}^{\beta/2}e^{i\omega_n \tau} d\tau=\delta_{n,0}\lim_{\beta\to\infty} \beta\,.
}
We should thus identify $\widetilde{\bold K}_{1/2,\ell,NCP}(\omega)$ with the quantity we get by first sending $\beta\to\infty$, then performing the $\tau$ integral, then throwing out any $\delta(\omega)$ that appear. 

As an illustrative example, let us begin by computing the $CP$-violating kernel $\widetilde{ K}^{\sigma\sigma}_{1/2,\ell,NCP}(\omega)$. Collecting the terms in \eqref{wickCPN} that include both $\hat G(x,x')$ and $ G(x,x')$ and plugging into the $\beta\to\infty$ limit of \eqref{fourierKernscal} we get
\es{DposNCP}{
\frac{1}{16\pi}\int_{-\infty}^{\infty} d\tau\int_0^\pi\sin\theta d\theta e^{i\omega\tau}P_\ell(\cos\theta)\cosh\tau\left[\frac{2\cosh\frac\tau2}{\sqrt{2\cosh\tau-2\cos\theta}}-1   \right]\,.
}
We then expand at large $\tau$ and perform the $\theta$ integral term by term to get
\es{DposCP}{
\frac{1}{16\pi(2\ell+1)}\int_{-\infty}^\infty d\tau e^{i\omega\tau}
\begin{cases}\sum_{p=\ell-1}^{\ell+2} \frac{e^{-p|\tau|}}{p}\,,\,\,\,\,\,\qquad \ell\geq2\\
1+\sum_{p=1}^{3} \frac{e^{-p|\tau|}}{p}\,,\qquad \ell=1\\
1+\frac{e^{-2|\tau|}}{2}\,,\qquad\qquad\,\,\,\, \ell=0\\
\end{cases}\,.
}
For $\ell=0,1$, the constant terms give delta functions, and so should be thrown out. Taking the Fourier transform of the remaining terms gives
\es{NCPD}{
\widetilde{K}^{\sigma\sigma}_{1/2,\ell,NCP}(\omega)=\frac{1}{8\pi(2\ell+1)}\sum_{p=\ell-1}^{\ell+2}\frac{p}{\omega^2+p^2}\,.
}
As a consistency check, if we send $\ell\to\infty$ this kernel vanishes as expected, as there is no analogous expression for $q=0$. We can perform a similar procedure for other entries in the symmetric matrix $\widetilde{\bold K}_{1/2,\ell,NCP}(\omega)$, which are all nonzero. We find similar answers, which are given in Appendix \ref{kernels}. Note that all the kernels are independent of $\sgn\kappa$, except for $\widetilde{ K}^{\sigma\mathcal{E}}_{1/2,\ell,NCP}(\omega)$ and $\widetilde{ K}^{\mathcal{E}\mathcal{B}}_{1/2,\ell,NCP}(\omega)$ which are proportional to $\sgn\kappa$.

\subsection{Numerical results}
\label{numerics}

Now that we have explicit expression for the relevant kernels in \eqref{q0D2}, \eqref{q0D2other}, \eqref{CSfourier2}, \eqref{CPD}, \eqref{NCPD}, \eqref{CPss}, and \eqref{NCP}, we can plug these values into \eqref{freeFinal} to compute the sub-leading scaling dimension $\Delta_{1/2}^{(1)}$ for $|\kappa|=1$. Note that when we take the eigenvalues of the kernel in \eqref{freeFinal}, the $\sgn\kappa$ factors in \eqref{NCP} cancel, so that we get the same expression $\Delta_{1/2}^{(1)}$ for $\kappa=\pm1$. Let us write this as 
\es{subScal}{
\Delta_{1/2}^{(1)}=\frac12\int\frac{d\omega}{2\pi}\sum_{\ell=0}^\infty(2\ell+1)L_{\ell}(\omega)\,.
}
As shown in Appendix \ref{asymp}, at large $\omega$ and $\ell$ the integrand in this expression behaves as
\es{largeL}{
L_\ell(\omega)=\left(\frac{2}{\omega^2+(\ell+1/2)^2}\right)\left(\frac{1}{1+\frac{64}{\pi^2}}\right)+\dots\,,
} 
which makes the integral \eqref{subScal} linearly divergent. The first factor in \eqref{largeL} equals the linear divergence that was found for $\kappa=0$ in \cite{Dyer:2015zha} if we set $\mu=q^2=1/4$ in that expression. As discussed in that case, there are several ways of regularizing this divergence. Using zeta functions, we can write
\es{zetaL}{
\left(\frac{1}{1+\frac{64}{\pi^2}}\right)\int\frac{d\omega}{4\pi}\sum_{\ell=0}^\infty\frac{2(2\ell+1)}{\omega^2+(\ell+1/2)^2}=\left(\frac{1}{1+\frac{64}{\pi^2}}\right)\sum_{\ell=0}^\infty\frac{2\ell+1}{2\ell+1}=\frac{\zeta(0,1/2)}{1+\frac{64}{\pi^2}}=0\,.
}
We can then subtract \eqref{zetaL} from \eqref{subScal} to get the expression
\es{subScalReg}{
\Delta_{1/2}^{(1)}=\frac12\int\frac{d\omega}{2\pi}\sum_{\ell=0}^\infty(2\ell+1)\left[L_{\ell}(\omega)-\left(\frac{2}{\omega^2+(\ell+1/2)^2}\right)\left(\frac{1}{1+\frac{64}{\pi^2}}\right)\right]\,,
}
which is no longer linearly divergent.

Another way of understanding \eqref{subScalReg} is that the inverse critical coupling $\mathfrak{g}^{-1}_c$, which we showed in Section \ref{leading} to vanish at leading order, obtains $1/N$ corrections. A similar phenomenon was encountered in \cite{Kaul:2008xw} when computing the thermal free energy of scalar QED$_3$ in flat space with $\kappa=0$ at sub-leading order in $1/N$, where the shift in the coupling was found to be $\delta \mathfrak{g}^{-1}_c=\frac{4}{N}\int\frac{d^3p}{(2\pi)^3}\frac{1}{p^2}$ for momentum $p$. In the large $N$ expansion, one effect of $\kappa\neq0$ is to shift the momentum $p^2\to p^2\left(1+\frac{64\kappa^2}{\pi^2}\right)$ \cite{Klebanov:2011td}, so that $\delta \mathfrak{g}^{-1}_c\to\delta \mathfrak{g}^{-1}_c\left(1+\frac{64\kappa^2}{\pi^2}\right)$. If we similarly shift the $\kappa=0$ expression computed in \cite{Dyer:2015zha} we get
\es{gcorrection}{
\frac{4\pi}{\mathfrak{g}_c}=\int\frac{d\omega}{2\pi}\sum_{\ell=0}^\infty(2\ell+1)\frac{1}{\omega^2+(\ell+1/2)^2}\left[1+\left(\frac{1}{1+\frac{64}{\pi^2}}\right)\frac4N\right]+O(1/N^2)\,.
} 
The $1/N$ term in this expression contributes to $F^{(1)}_q$ through the second term in \eqref{free0} precisely as the subtraction implemented in \eqref{subScalReg}.

Even after the linear divergence has been regularized, there is still a potential logarithmic divergence in the integral in \eqref{subScalReg}. This logarithmic divergence cancels as long as we regularize this integral consistent with conformal symmetry, such as by using the symmetric cutoff
\es{cutoff}{
(\ell+1/2)^2+\omega^2<\Lambda^2\,.
}
This can be thought of as preserving rotational invariance on $\mathbb{R}^3$, as the high energy modes are insensitive to the curvature of the sphere. The cancellation of this divergence is a strong check on our results, as the potentially logarithmic divergence in the integrand of \eqref{subScalReg} receives contributions from all the kernels evaluated in the previous sections. With this cutoff, we can now numerically perform the sum and integral in \eqref{subScalReg} for large $\Lambda$, and then extrapolate to $\Lambda$. We find that the numerics rapidly converge, and yield the final answer
\es{FINAL}{
\Delta_{1/2}^{(1)}\approx-0.2789\,.
}

\section{Energy splitting}
\label{splitting}

The sub-leading correction to the scaling dimension given in \eqref{FINAL} is expected to apply to monopoles in different representations of the symmetry group $SU(N)\times U(1)_\text{top}$ with different Lorentz spins $\ell$. As shown in \cite{Chester:2017vdh} for the case $q=1/2$ and $|\kappa|=1$, this degeneracy applies to monopoles with spins $\ell<O(\sqrt{N})$ and the associated $SU(N)$ representation
 \es{RellScalar}{
  {\bf R}_{\ell} \equiv    \raisebox{-0.15in}{\makebox[0in][l]{$\underbrace{\rule{0.68in}{0in}}_{N/2 - \ell}$}}
   \raisebox{0.25in}{\makebox[0in][l]{$\overbrace{\rule{1.2in}{0in}}^{N/2 + \ell}$}}
   \begin{ytableau}
 {}& {} & \none[\cdots] & {} & {} & \none[\cdots] & {} \\
 {}& {} & \none[\cdots] & {} 
\end{ytableau}\,,  \quad
 \dim {\bf R}_\ell =  \frac{(2 \ell + 1)}{N - 1} \begin{pmatrix} 3N/2 + \ell - 1\\ N \xi + \ell + 1 \end{pmatrix}
  \begin{pmatrix} 3N/2 - \ell - 2 \\ N \xi - \ell \end{pmatrix} \,.
 }  
 When $\ell=O(\sqrt{N})$, a formula for the energy splitting term in \eqref{largeB} was also derived based on the $\frac{\log\beta}{\beta}$ terms that appear in \eqref{largeB}, which in our case yields
 \es{Esplit}{
 \frac{\delta E_\ell}{N}=\frac{\ell^2}{N} \left(\frac{3}{16\sqrt{2\pi}}\right)\begin{pmatrix}
   1 &   2 & -i
   \end{pmatrix} \widetilde{\bold K}^{-1}_{1/2,1}(0)
   \begin{pmatrix}
    1 \\   2 \\ -i
   \end{pmatrix} \,.
 }
 Since the energy splitting goes like $\ell^2/N$, we see that the large $N$ expansion starts to break down when $\ell=O(\sqrt{N})$. Using the expression for $\widetilde{\bold K}_{1/2,1}(0)$ calculated in the previous sections, we find
\es{splitFINAL}{
 \frac{\delta E}{N}=-\frac{-25920 + 2052 \pi^2 + 63 \pi^4}{-35328 + 10048 \pi^2 - 10 \pi^4 + 18 \pi^6}\frac{\ell^2}{N}\approx-.0059\frac{\ell^2}{N}\,.
}

\section{Conclusion}
\label{conc}

The main result of this paper is that the scaling dimension of $q=1/2$ monopoles with spin $\ell<O(\sqrt{N})$ and $SU(N)$ irrep \eqref{RellScalar} in scalar QED$_3$ with $|\kappa|=1$ and $N$ scalar flavors is
\es{finalDim}{
\Delta_{1/2}=N-0.2789+O(1/N)\,,
}
which is obtained by combining the leading order result \cite{Chester:2017vdh} given in \eqref{SpecialCase} with \eqref{FINAL}. The $O(1)$ correction to $\Delta_{1/2}$ was obtained by performing a Gaussian integral over the fluctuations of the gauge field $A_\mu$ and the Lagrange multiplier $\lambda$ on $S^2 \times S^1_\beta$ with $2\pi$ magnetic flux as $\beta\to\infty$. The scaling dimension \eqref{finalDim} does not depend on the sign of $\kappa$, which is also the case in the large $N$ computation of non-monopole operators in Abelian gauge theories \cite{Gracey:1992zc}.\footnote{For non-monopole operators, the effect of $\kappa\neq0$ to sub-leading order in large $N$ is just to shift $N\to N\sqrt{1+\frac{64\kappa^2}{\pi^2}}$ \cite{Gracey:1992zc,Klebanov:2011td}.}

When $\ell=O(\sqrt{N})$, the sub-leading term is affected by a spin-dependent contribution given in \eqref{splitFINAL}. The coefficient $-.0059$ of this splitting term is much smaller than the coefficients in \eqref{finalDim}, so we expect monopoles to be highly degenerate in the regime where these formulae apply. In \cite{Chester:2017vdh}, a spin-dependent term analogous to \eqref{splitFINAL} was found for QED$_3$ with $\kappa=0$ and $q=1$, except with the positive and much large coefficient $.145$. It would be interesting to understand the physical meaning of the large discrepancy between the splitting in these two cases, and any implications this has for the dualities in \cite{Seiberg:2016gmd,Komargodski:2017keh}.

A striking feature of the computation in this work is that the scalar thermal Green's function on $S^2\times S^1_\beta$ with $4\pi q$ magnetic flux and $|\kappa|=q+1/2$ can be written in closed form, which allowed us to find simple analytical expressions for all the integration kernels. This simplification occurred because the $\lambda$ saddle point value $\mu$ exactly cancelled the contribution $-q^2$ of the magnetic flux in the eigenvalue $\lambda_\ell$ \eqref{spectrum} of the Klein-Gordon operator in this case. The same simplification also occurs for BPS monopoles in $\mathcal{N}=2$ SQED$_3$, because of the Lagrange multiplier fields in that case. For this theory, the scaling dimension can be found exactly for any $q$ and $\kappa$ using supersymmetric localization and $F$-maximization \cite{Benini:2011cma,Benini:2009qs,Klebanov:2011td}. The localization result can be expanded in large $N$, e.g. for $q=1/2$ and $\kappa=0$, to get
\es{largeNBPS}{
\Delta^{\mathcal{N}=2}_\text{BPS}=\frac{N}{2}+\frac{2}{\pi^2}+O(1/N)\,.
}
If we enhance the supersymmetry to $\mathcal{N}=4$, then $\Delta^{\mathcal{N}=4}_\text{BPS}=\frac{N}{2}$ exactly. It would be interesting to compute the sub-leading corrections in these cases using the methods of this paper, to see how the simplified Green's function leads to such simple analytic formulae. 

Finally, it would be interesting to compare the large $N$ calculation in this work to other non-perturbative numerical methods. In the $k=0$ scalar QED$_3$ case, the subleading in $1/N$ calculation of \cite{Dyer:2015zha} was found to match finite $N$ quantum Monte Carlo studies in \cite{2009PhRvB..80r0414L,2012PhRvL.108m7201K,2013PhRvL.111m7202B} even down to $N=2$. For $k=0$ fermionic QED$_3$, the Monte Carlo estimates in \cite{Karthik:2018rcg,Karthik:2019mrr} match the subleading in $1/N$ estimate in \cite{Pufu:2013vpa} for large $N$, but differ at small $N$. It would be interesting to perform a similar Monte Carlo study for scalar QED$_3$ with $k\neq0$, and see how the results compare to the subleading in $1/N$ prediction in this work. Such a study could also be used to verify the $N=1$ duality of \cite{Seiberg:2016gmd}, which predicts that $\Delta_{1/2}=1$ exactly.

\subsection*{Acknowledgments}

I thank Ofer Aharony, Silviu Pufu, Mark Mezei, Luca Iliesiu, and Nathan Agmon for helpful discussions, Ofer Aharony for reviewing the draft, and Eric Dupuis and William Witczak-Krempa for pointing out an error in the original version. I am supported by the Zuckerman STEM Leadership Fellowship.

\appendix
\section{Matter kernel formulae}
\label{kernels}

In sections \ref{CP} and \ref{noCP} we explicitly evaluated the $CP$-preserving and $CP$-violating components of the $\widetilde{K}^{\sigma\sigma}_{1/2,\ell}(\omega)$ kernel, respectively, which we denote as $\widetilde{K}^{\sigma\sigma}_{1/2,\ell,CP}(\omega)$ and $\widetilde{K}^{\sigma\sigma,NCP}_{1/2,\ell}(\omega)$. In this appendix we give explicit expression for the other $CP$-violating and $CP$-preserving components of the matter kernels.

For the $CP$-preserving kernels, we have
\es{CPss}{
\widetilde{K}^{\mathcal{B}\mathcal{B}}_{1/2,\ell,CP}(\omega)=&-\frac{1}{4\pi} -\frac{\ell^2}{{\ell(\ell+1)}}D_\ell(\omega) +\frac{(-1)^{\ell}}{2\pi{\ell(\ell+1)}}\sum_{p=1}^\ell(-1)^{p}\frac{\left(\ell(\ell+1)-p^2\right)^2}{p^2+\omega^2}\\
&+\frac{(-1)^\ell{\ell(\ell+1)}}{4\pi\omega^2}\left(1-\pi\omega\coth(\pi\omega)\right)+\frac{\omega}{16}\tanh(\pi\omega/2)^{(-1)^\ell}\\
&+\frac14D_0(\omega)+  \frac{(-1)^\ell }{\ell(\ell+1)}  \left[1-\frac74\ell+\ell^2/4+4\ell^3+2\ell^4\right]D_0(\omega)   \\
&+\sum_{p=1}^{\ell-1}(-1)^{\ell+p}\frac{2p+1}{4\ell(\ell+1)}\left[4+6p(p+1)-\ell(\ell+1)(7-(-1)^{p+\ell})\right]D_p(\omega)\,,\\
\widetilde{K}^{\mathcal{E}\mathcal{E}}_{1/2,\ell,CP}(\omega)=&\left(1+\frac{\omega^2}{\ell(\ell+1)}\right)\left[\frac{\ell(\ell+1)}{2}D_\ell(\omega)-\sum_{p=0}^\ell\frac{(-1)^{p+\ell}p^2}{2\pi(p^2+\omega^2)}\right.\\
&+\left.\sum_{p=0}^{\lceil \ell/2 \rceil-1}\left((2\ell-4p-1)D_{\ell-1-2p}(\omega)-{(\ell-1-2p)D_{\ell-2-2p}(\omega)}-{(\ell-2p)D_{\ell-2p}(\omega)}\right)\right]\,,\\
\widetilde{K}^{\sigma\mathcal{B}}_{1/2,\ell,CP}(\omega)=&i(-1)^\ell{\frac{  1  }{\sqrt{\ell(\ell+1)}}}\left[(1-2\ell^2-2\ell)D_0(\omega) +\sum_{p=1}^{\ell-1}(-1)^{p}(2p+1)D_p(\omega)  \right.\\
& \left.+(-1)^\ell \ell D_\ell(\omega)-\frac{{\ell(\ell+1)}}{4\pi\omega^2}\left(1-\pi\omega\coth(\pi\omega)\right)-\frac{1}{2\pi}\sum_{p=1}^\ell(-1)^{p}\frac{\ell(\ell+1)-p^2}{p^2+\omega^2}\right]\,,\\
}
where $D_\ell(\omega)$ is defined in \eqref{KDfree}. 

For the $CP$-violating kernels, we have
\es{NCP}{
\widetilde{K}^{{\sigma}\mathcal{B}}_{1/2,\ell,NCP}(\omega)&=\frac{i}{8\pi(2\ell+1)\sqrt{\ell(\ell+1)}}\left[  -\frac{\ell^2-1}{(\ell-1)^2+\omega^2}+\frac{\ell^2}{\ell^2+\omega^2}\right.\\
&\left.\qquad\qquad\qquad\qquad\qquad \qquad-\frac{(\ell+1)^2}{(\ell+1)^2+\omega^2} +\frac{\ell(\ell+2)}{(\ell+2)^2+\omega^2}  \right]\,,\\
\widetilde{K}^{\sigma\mathcal{E}}_{1/2,\ell,NCP}(\omega)&=\frac{\sgn\kappa}{8\pi(2\ell+1)}\sqrt{1+\frac{\omega^2}{\ell(\ell+1)}}\left[   \frac{\ell^2-1}{(\ell-1)^2+\omega^2}+\frac{\ell(\ell+2)}{\ell^2+\omega^2}\right.\\
&\left.\qquad\qquad\qquad\qquad\qquad \qquad\quad-\frac{\ell^2-1}{(\ell+1)^2+\omega^2} -\frac{\ell(\ell+2)}{(\ell+2)^2+\omega^2}  \right]\,,\\
\widetilde{K}^{\mathcal{E}\mathcal{E}}_{1/2,\ell,NCP}(\omega)&=\frac{1}{8\pi}\left(1+\frac{\omega^2}{\ell(\ell+1)}\right)\left[
\frac{1}{(2\ell+1)}\left(\frac{(\ell-1)(\ell+1)^2}{(\ell-1)^2+\omega^2}+\frac{\ell(\ell+2)^2}{\ell^2+\omega^2}\right.\right.\\
&\left.\left.\qquad\qquad\qquad\qquad\qquad+\frac{(\ell-1)^2(\ell+1)}{(\ell+1)^2+\omega^2}+\frac{\ell^2(\ell+2)}{(\ell+2)^2+\omega^2}\right)\right]\,,\\
\widetilde{K}^{\mathcal{B}\mathcal{B}}_{1/2,\ell,NCP}(\omega)&=
\frac{1}{4\pi}-\frac{1}{8\pi(2\ell+1)\ell(\ell+1)}\left[\frac{(\ell-1)(\ell+1)^2}{(\ell-1)^2+\omega^2}+\frac{\ell^3}{\ell^2+\omega^2}\right.\\
&\qquad\qquad\qquad\qquad\qquad\qquad\quad\left.+\frac{(\ell+1)^3}{(\ell+1)^2+\omega^2}+\frac{\ell^2(\ell+2)}{(\ell+2)^2+\omega^2}\right]\,,\\
\widetilde{K}^{\mathcal{E}\mathcal{B}}_{1/2,\ell,NCP}(\omega)&=\sgn(\kappa)
\frac{i\sqrt{\ell(\ell+1)+\omega^2}}{8\pi(2\ell+1)\ell(\ell+1)}\left[-\frac{(\ell-1)(\ell+1)^2}{(\ell-1)^2+\omega^2}+\frac{\ell^2(\ell+2)}{\ell^2+\omega^2}\right.\\
&\qquad\qquad\qquad\qquad\qquad\qquad\quad\left.+\frac{(\ell-1)(\ell+1)^2}{(\ell+1)^2+\omega^2}-\frac{\ell^2(\ell+2)}{(\ell+2)^2+\omega^2}\right]\,.
}

\section{Comparison to \cite{Dyer:2015zha}}
\label{conventions}

In this appendix, we will relate the $CP$-preserving kernels $\widetilde{\bold K}_{1/2,\ell,CP}(\omega)$ computed in Section \ref{CP} to the results of $\kappa=0$ kernels computed in \cite{Dyer:2015zha}, which we use to justify our choice of regularization for $\widetilde{K}^{\mathcal{E}\mathcal{E}}_{1/2,\ell,CP}(\omega)$ and $\widetilde{K}^{\mathcal{B}\mathcal{E}}_{1/2,\ell,CP}(\omega)$.

In \cite{Dyer:2015zha}, the matter kernels were computed on $S^2\times \mathbb{R}$ by expanding the gauge fluctuation in the gauge redundant basis
 \es{aFourierS}{
  a(x) &= \int\frac{d\omega}{2\pi} \left[\mathfrak{a}_{00}^E(0) \frac{d\tau}{\sqrt{4 \pi }}  + 
 \sum_{\ell=1}^\infty \sum_{m=-\ell}^\ell 
   \left[   \mathfrak{a}_{\ell m}^\tau(\omega) Y_{\ell m}(x)d\tau+ \mathfrak{a}_{\ell m}^E(\omega) \frac{dY(x)_{\ell m}}{\sqrt{\ell(\ell+1)}}\right.\right.\\
&\left.\left.   \qquad\qquad\qquad\qquad\qquad \qquad\qquad\qquad\qquad+ \mathfrak{a}_{\ell m}^B(\omega)\frac{*_2dY(x)_{\ell m}}{\sqrt{\ell(\ell+1)}}\right] \right]{e^{-i \omega \tau}}\,,\\
 }
 so that the matrix of frequency space kernels analogous to \eqref{freeFinal} is a $4\times4$ matrix
 \es{matrixS}{
 \begin{pmatrix}
 \widetilde{K}^{\sigma\sigma}_{q,\ell}(\omega) & \widetilde{K}^{\sigma B}_{q,\ell}(\omega)& \widetilde{K}^{\sigma\tau}_{q,\ell}(\omega)& \widetilde{K}^{\sigma E}_{q,\ell}(\omega)\\
 -\widetilde{K}^{  \sigma B* }_{q,\ell}(\omega) & \widetilde{K}^{B B}_{q,\ell}(\omega)& \widetilde{K}^{ \tau B }_{q,\ell}(\omega)& \widetilde{K}^{ BE}_{q,\ell}(\omega)\\
  \widetilde{K}^{  \sigma\tau * }_{q,\ell}(\omega) & -\widetilde{K}^{ \tau B * }_{q,\ell}(\omega)& \widetilde{K}^{\tau\tau}_{q,\ell}(\omega)& \widetilde{K}^{\tau E}_{q,\ell}(\omega)\\
   \widetilde{K}^{ \sigma E*}_{q,\ell}(\omega) &- \widetilde{K}^{BE*}_{q,\ell}(\omega)& \widetilde{K}^{\tau E *}_{q,\ell}(\omega)& \widetilde{K}^{E E}_{q,\ell}(\omega)\\
 \end{pmatrix}\,,
 }
 where the signs are determined by the reality of the position space kernels \eqref{wickCPN}. Since the basis in \eqref{aFourierS} is gauge redundant, \eqref{matrixS} has a zero eigenvector that relates the entries as 
\es{eigenvector}{
\widetilde{K}_{\ell}^{q,\sigma E}(\omega)=&\frac{-i\omega}{\sqrt{\ell(\ell+1)}}\widetilde{K}_{\ell}^{q,\sigma \tau}(\omega)\,,\qquad
  \widetilde{K}_{\ell}^{q,B E}(\omega)=\frac{-i\omega}{\sqrt{\ell(\ell+1)}}\widetilde{ K}_{\ell}^{q,B \tau}(\omega)\,,\\
\widetilde{K}_{\ell}^{q,\tau E}(\omega)=&\frac{-i\omega}{\sqrt{\ell(\ell+1)}}\widetilde{K}_{\ell}^{q,\tau \tau}(\omega)\,,\qquad
\widetilde{  K}_{\ell}^{q,E E}(\omega)=\frac{\omega^2}{{\ell(\ell+1)}}\widetilde{K}_{\ell}^{q,\tau \tau}(\omega)\,.\\
}
By comparing \eqref{aFourierS} and \eqref{aFourier}, we see that the different independent entries in \eqref{matrixS} are related to our $\widetilde{\bold K}_{q,\ell}(\omega)$ as
\es{relation}{
\widetilde{K}^{\mathcal{B}\mathcal{B}}_{q,\ell}(\omega)&=\widetilde{K}^{{B}{B}}_{q,\ell}(\omega)\,,\qquad\qquad\qquad\qquad
 \widetilde{K}^{\mathcal{E}\mathcal{E}}_{q,\ell}(\omega)=\left(1+\frac{\omega^2}{\ell(\ell+1)}\right)\widetilde{K}^{{\tau}{\tau}}_{q,\ell}(\omega)\,,\\
  \widetilde{K}^{{\sigma}\mathcal{E}}_{q,\ell}(\omega)&=\sqrt{1+\frac{\omega^2}{\ell(\ell+1)}}\widetilde{K}^{{\sigma}{\tau}}_{q,\ell}(\omega)\,,\qquad \widetilde{K}^{\mathcal{B}\mathcal{E}}_{q,\ell}(\omega)=\sqrt{1+\frac{\omega^2}{\ell(\ell+1)}}\widetilde{K}^{{\tau}{B}}_{q,\ell}(\omega)\,,\\
  \widetilde{K}^{\mathcal{B}\sigma}_{q,\ell}(\omega)&=\widetilde{K}^{\sigma\mathcal{B}}_{q,\ell}(\omega)=\widetilde{K}^{\sigma B}_{q,\ell}(\omega)\,.\\
}

The kernels in \eqref{matrixS} were computed in \cite{Dyer:2015zha} for $\kappa=0$, in which case the theory is $CP$-invariant so $ \widetilde{K}^{\sigma\tau}_{q,\ell}(\omega)$, $ \widetilde{K}^{\sigma E}_{q,\ell}(\omega)$, $ \widetilde{K}^{\tau B}_{q,\ell}(\omega)$, and $ \widetilde{K}^{BE}_{q,\ell}(\omega)$ vanish. For $q=1/2$ and for arbitrary $\mu$, we will denote the remaining kernels as $\mathcal{K}_{1/2,\ell}(\omega)$, and they are given by the following infinite sums:
\es{sKernels}{
\mathcal{K}_{1/2,\ell}^{\sigma\sigma}(\omega)=&\frac{8\pi^2}{2\ell+1}\sum_{\ell',\ell''=1/2}^{\infty}\frac{\lambda_{\ell'}+\lambda_{\ell''}}{2\lambda_{\ell'}\lambda_{\ell''}(\omega^2+(\lambda_{\ell'}+\lambda_{\ell''})^2)}\mathcal{I}_D(\ell,\ell',\ell'')\,,\\
\mathcal{K}^{\sigma B}_{1/2,\ell}(\omega)=&\frac{8\pi^2 i}{(2\ell+1)\sqrt{\ell(\ell+1)}}\sum_{\ell',\ell''=1/2}^{\infty}\frac{\lambda_{\ell'}+\lambda_{\ell''}}{2\lambda_{\ell'}\lambda_{\ell''}(\omega^2+(\lambda_{\ell'}+\lambda_{\ell''})^2)}\mathcal{I}_H(\ell,\ell',\ell'')\,,\\
\mathcal{K}^{\tau E}_{1/2,\ell}(\omega)=&\frac{8\pi^2 i\omega}{(2\ell+1)\sqrt{\ell(\ell+1)}}\sum_{\ell',\ell''=1/2}^{\infty}\frac{(\lambda_{\ell'}-\lambda_{\ell''})\left[\ell''(\ell''+1)-\ell'(\ell'+1)\right]}{2\lambda_{\ell'}\lambda_{\ell''}(\omega^2+(\lambda_{\ell'}+\lambda_{\ell''})^2)}\mathcal{I}_D(\ell,\ell',\ell'')\,,\\
\mathcal{K}^{EE}_{1/2,\ell}(\omega)=&\frac{-8\pi^2}{(2\ell+1)\ell(\ell+1)}\sum_{\ell'=1/2}^{\infty}\left[\sum_{\ell''=1/2}^\infty\frac{(\lambda_{\ell'}+\lambda_{\ell''})\left[\ell'(\ell'+1)-\ell''(\ell''+1)\right]^2}{2\lambda_{\ell'}\lambda_{\ell''}(\omega^2+(\lambda_{\ell'}+\lambda_{\ell''})^2)}\mathcal{I}_D(\ell,\ell',\ell'') \right.\\
&\left.\qquad\qquad\qquad\qquad\qquad\qquad\qquad\qquad\qquad\qquad\qquad\qquad+\frac{2\ell'+1}{8\pi \lambda_{\ell'}}\right]+C_q\,,\\
\mathcal{K}^{BB}_{1/2,\ell}(\omega)=&\frac{8\pi^2}{(2\ell+1)\ell(\ell+1)}\sum_{\ell'=1/2}^{\infty}\left[\sum_{\ell''=1/2}^\infty\frac{\lambda_{\ell'}+\lambda_{\ell''}}{2\lambda_{\ell'}\lambda_{\ell''}(\omega^2+(\lambda_{\ell'}+\lambda_{\ell''})^2)}\mathcal{I}_B(\ell,\ell',\ell'')
+\frac{2\ell'+1}{8\pi \lambda_{\ell'}}\right]+C_q\,,\\
\mathcal{K}^{\tau\tau}_{1/2,\ell}(\omega)=&\frac{8\pi^2}{(2\ell+1)}\sum_{\ell'=1/2}^{\infty}\left[\sum_{\ell''=1/2}^\infty\frac{-(\lambda_{\ell'}+\lambda_{\ell''})(\omega^2+4\lambda_{\ell'}\lambda_{\ell''})}{2\lambda_{\ell'}\lambda_{\ell''}(\omega^2+(\lambda_{\ell'}+\lambda_{\ell''})^2)}\mathcal{I}_D(\ell,\ell',\ell'')+\frac{2\ell'+1}{4\pi \lambda_{\ell'}}\right]\,,\\
}
where
\es{Is}{
\mathcal{I}_D(\ell,\ell',\ell'')=&\left[\frac{(2\ell+1)(2\ell'+1)(2\ell''+1)}{32\pi^3}\right]\Bigg(\begin{matrix}\ell&\ell'&\ell''\\[-12pt]0&-1/2&1/2\end{matrix}\Bigg)^2\,,\\
\mathcal{I}_H(\ell,\ell',\ell'')=& \mathcal{I}_D(\ell,\ell',\ell'')\times \begin{cases} 
\ell(\ell+1)-(\ell'-\ell'')^2\,\,\,\,\quad\quad\text{for $\ell+\ell'+\ell''$ odd} \\
\ell(\ell+1)-(\ell'+\ell''+1)^2\quad\text{for $\ell+\ell'+\ell''$ even} \\
 \end{cases}  \\
\mathcal{I}_B(\ell,\ell',\ell'')=&- \mathcal{I}_D(\ell,\ell',\ell'')\times \begin{cases} 
\left[(\ell'-\ell'')^2-\ell(\ell+1)\right]^2\,\,\,\,\quad\quad\text{for $\ell+\ell'+\ell''$ odd} \\
\left[(\ell'+\ell''+1)^2-\ell(\ell+1)\right]^2\quad\text{for $\ell+\ell'+\ell''$ even.} \\
 \end{cases}  \\\
}
The constants $C_q$ was defined by the regularization used in \cite{Dyer:2015zha} as
 \es{CqDef}{
  C_q \equiv 
    \frac{1}{4 \pi} \left[ \sum_{j' = q}^{\infty} \left( \frac{j' +1/2}{\sqrt{(j' + 1/2)^2 + \mu_q^2 - q^2}} - 1 \right) - q \right] \,.
 }
While this quantity was zero for the $k=0$ theory, for general $k$ we see from saddle point equation \eqref{muSad} that it is
\es{Cqus}{
C_q=-\frac{q|\kappa|}{4\pi\lambda_q}=-\frac{1}{8\pi}\,,
}
where in the second equality we specified to $|\kappa|=q+1/2$ and $q=1/2$. Since both $\mathcal{K}_{1/2,\ell}(\omega)$ and our $\widetilde{\bold K}_{1/2,\ell,CP}(\omega)$ are independent of $\kappa$, we can relate them by setting $\mu=q^2=1/4$ in $\mathcal{K}_{1/2,\ell}(\omega)$. We can check numerically that they are equal, which justifies our choice of regularization, and is a check on our computation of these terms in general.

\section{Asymptotic expansion}
\label{asymp}

In this appendix we will write the asymptotic formula at large $\ell$ and $\omega$ for the integrand in \eqref{freeFinal}. 

The expression for the $CP$-violating kernels in \eqref{NCP} and \eqref{NCPD}, the Chern-Simons kernel in \eqref{CSfourier2}, and the $q=0$ kernels in \eqref{q0D2other} and \eqref{KDfree} are all simple functions of $\ell$ and $\omega$ whose asymptotic expansion is straightforward. The $CP$-preserving kernels in \eqref{CPss} and \eqref{CPD} include sums of the form $\sum_{p=1}^\ell D_p(\omega)$ and $\sum_{p=1}^\ell \frac{f(p)}{p^2+\omega^2}$, for some polynomial $f(p)$. The former can be written in terms of gamma functions using standard identities, while the latter can be expressed in terms of the polygamma function $\psi(z)$, both of which have standard asymptotic formulae. The resulting asymptotic expressions for the kernels are
\es{ass}{
\widetilde{K}^{{\sigma}\sigma}_{1/2,\ell}(\omega)&=\widetilde{K}^{{\sigma}\sigma}_{0,\ell}(\omega)+\frac{(\ell+\frac12)^2-2\omega^2}{2\pi\left[(\ell+\frac12)^2+\omega^2\right]^{3}}+\frac{2(\ell+\frac12)^4+{25}\omega^4-{45}\omega^2(\ell+\frac12)^2}{8\pi\left[(\ell+\frac12)^2+\omega^2\right]^5}+\dots \,, \\
\widetilde{K}^{{\sigma}\mathcal{B}}_{1/2,\ell}(\omega)&=i\frac{\ell+\frac12}{16\left[(\ell+\frac12)^2+\omega^2\right]^{3/2}}+i\frac{2(\ell+\frac12)^4+\omega^4-17\omega^2(\ell+\frac12)^2}{128(\ell+\frac12)\left[(\ell+\frac12)^2+\omega^2\right]^{7/2}}+\dots \,, \\
\widetilde{K}^{\mathcal{B}\mathcal{B}}_{1/2,\ell}(\omega)&=\widetilde{K}^{\mathcal{B}\mathcal{B}}_{0,\ell}(\omega)+\frac{1}{16\sqrt{(\ell+\frac12)^2+\omega^2}}-\frac{1}{2\pi\left[(\ell+\frac12)^2+\omega^2\right]}+\dots\,,  \\
\widetilde{K}^{\mathcal{E}\mathcal{E}}_{1/2,\ell}(\omega)&=\widetilde{K}^{\mathcal{E}\mathcal{E}}_{0,\ell}(\omega)+\left(\ell(\ell+1)+\omega^2)\right)\left[\frac{1}{16\left[{(\ell+\frac12)^2+\omega^2}\right]^{3/2}}+\frac{2(\ell+\frac12)^2-\omega^2}{2\pi\left[(\ell+\frac12)^2+\omega^2\right]^3}\right]+\dots\,,  \\
\widetilde{K}^{\mathcal{E}\sigma}_{1/2,\ell}(\omega)&=\sgn(\kappa)\sqrt{1+\frac{\omega^2}{\ell(\ell+1)}}\left[\frac{(\ell+\frac12)^2}{2\pi\left[{(\ell+\frac12)^2+\omega^2}\right]^{2}}+\frac{5(\ell+\frac12)^4-\omega^4-24\omega^2(\ell+\frac12)^2}{8\pi\left[(\ell+\frac12)^2+\omega^2\right]^4}\right]+\dots\,,  \\
\widetilde{K}^{\mathcal{E}\mathcal{B}}_{1/2,\ell}(\omega)&=-\sgn(\kappa)i\sqrt{\ell(\ell+1)+{\omega^2}}\left[-\frac{1}{2\pi}+\frac{\left((\ell+\frac12)^2-\omega^2\right)}{2\pi\left[{(\ell+\frac12)^2+\omega^2}\right]^{3}}\right.\\
&\left.\qquad\qquad\qquad\qquad\qquad+\frac{19(\ell+\frac12)^6+\omega^6-133\omega^2(\ell+\frac12)^4+9\omega^4(\ell+\frac12)^2}{16\pi(\ell+\frac12)^2\left[(\ell+\frac12)^2+\omega^2\right]^5}\right]+\dots\,,  \\
}
where the $q=0$ kernels are all simple functions of
\es{Doasymp}{
\widetilde{K}^{{\sigma}\sigma}_{0,\ell}(\omega)=\frac{1}{8\sqrt{(\ell+\frac12)^2+\omega^2}}+\frac{-(\ell+\frac12)^2+\omega^2}{64\left[(\ell+\frac12)^2+\omega^2\right]^{5/2}}+\dots\,.
}
We can now plug these expressions into \eqref{subScal} to get
\es{largeL2}{
L_\ell(\omega)=\frac{2\pi^2}{(\pi^2+64)\left[(\ell+\frac12)^2+\omega^2\right]}+\frac{4(64+3\pi^2)((\ell+\frac12)^2-2\omega^2)}{\pi(64+\pi^2)\left[(\ell+\frac12)^2+\omega^2\right]^{5/2}}+O\left(\frac{1}{[(\ell+\frac12)^2+\omega^2]^2}\right)\,,
} 
The first term is the linear divergence that we regularize in \eqref{zetaL}. The second term gives a potential logarithmic divergence, that cancels in the sum and integral in \eqref{subScal} as long as we use a regulator that respects conformal invariance, such as \eqref{cutoff}. The higher order terms are all convergent.

\bibliographystyle{ssg}
\bibliography{CSsub}

\end{document}